\documentstyle[12pt]{article}

\begin{document}
\title{$\Upsilon(1s)\rightarrow\gamma f_2(1270)$ Decay}
\author{Bing An Li\\
Department of Physics and Astronomy, University of Kentucky\\
Lexington, KY 40506, USA}

\maketitle
\begin{abstract}
Decay $\Upsilon(1s)\rightarrow\gamma f_2(1270)$ is studied by an approach in which 
the tensor meson, $f_2(1270)$, is strongly coupled to gluons. Besides the strong suppression of
the amplitude $\Upsilon(1s)\rightarrow\gamma gg, gg\rightarrow f_2$ by the mass of b-quark, 
d-wave dominance in $\Upsilon(1s)\rightarrow\gamma f_2(1270)$ is revealed from this approach, which  
provides a large enhancement. The combination of these two 
factors leads to larger $B(\Upsilon(1s)\rightarrow\gamma f_2(1270))$.
The decay rate of $\Upsilon(1s)\rightarrow\gamma f_2(1270)$ and the ratios of 
the helicity amplitudes are obtained and they are in agreement with data.

\end{abstract}
\newpage

The measurements 
\begin{eqnarray}
\lefteqn{B(\Upsilon(1S)\rightarrow\gamma f_2(1270))=(10.2\pm0.8\pm0.7)\times 10^{-5},}\\
&&B(\Upsilon(1S)\rightarrow\gamma f_2(1270))=(10.5\pm1.6(stat)^{+1.9}_{-1.8}(syst))\times 10^{-5} 
\end{eqnarray}
have been reported by CLEO in the channel of $\Upsilon(1S)\rightarrow\gamma  
f_2(1270), f_2(1270)\rightarrow\pi^+\pi^-$[1]
and $f_2\rightarrow\pi^0\pi^0$[2] respectively. It is known that
\begin{equation}
B(J/\psi\rightarrow\gamma f_2(1270)=(1.43\pm0.11)\times 10^{-3}[3].
\end{equation}
$B(\Upsilon(1S)\rightarrow\gamma f_2(1270))$ is about one order of magnitude smaller than 
$B(J/\psi\rightarrow\gamma f_2(1270))$. CLEO Collaboration has reported the measurements of
$B(\Upsilon(1S)\rightarrow\gamma\eta'(\eta))$ whose upper limits are smaller than $B(J/\psi\rightarrow\gamma\eta'(\eta))$
by almost three order of magnitudes[4].
In Refs.[5] the dependencies of $B(J/\psi,\Upsilon(1S)\rightarrow\gamma\eta'(\eta))$ on corresponding
quark masses are found and explanation of very small $B(\Upsilon\rightarrow\gamma\; (\eta', \eta))$ is presented. 
The question is that comparing with $B(J/\psi,\Upsilon(1S)\rightarrow\gamma\eta'(\eta))$,
why $B(\Upsilon\rightarrow\gamma\;f_2)$ is not too small.
$B(\Upsilon\rightarrow\gamma\;f_2)$ has been studied by many authors.
In Ref.[6] a QCD analysis for $B(\Upsilon(1S)\rightarrow\gamma f_2(1270))$ has been done. 
In Ref.[7] the ratio $\frac{B(\Upsilon\rightarrow\gamma f_2)}{B(J/\psi\rightarrow\gamma f_2)}$ has been studied by using 
soft-collinear theory and nonrelativistic QCD.
In 1983 we have studied the radiative decay $J/\psi\rightarrow\gamma f_2(1270)$[8].
In this letter the same approach exploited in Ref.[8] is used to study  
$\Upsilon\rightarrow\gamma f_2$. 

The study done in Ref.[8] is based on the arguments presented in Refs.[9] that the tensor meson $f_2(1270)$ contains
glueball components
\begin{equation}
|f_2>=cos \phi |q\bar{q}>+sin\phi |gg>.
\end{equation}
Tensor glueball has been studied by many authors[10]. Lattice QCD predicts the existence of light $2^{++}$ glueball[11].
It is reasonable to assume that there is mixing between $f_2(1270)$ and a tensor glueball. 
In radiative decay $J/\psi\rightarrow\gamma f_2$ the $q\bar{q}$ component of $f_2(1270)$
is suppressed by $O(\alpha^2_s(m_c))$[9]
\begin{equation}
\frac{\Gamma(J/\psi\rightarrow\gamma+(q\bar{q}))}{\Gamma(J/\psi\rightarrow\gamma+(gg))}\sim \alpha^2_s(m_c).
\end{equation}
Therefore, the glueball component of $f_2$ is
dominant in the decay $J/\psi\rightarrow\gamma f_2$. It is the same that 
the glueball component of $f_2$ is
dominant in the decay $\Upsilon(1S)\rightarrow\gamma f_2$ too.
In QCD the radiative decays $J/\psi, \Upsilon\rightarrow\gamma\; f_2$ are described
as $J/\psi, \Upsilon\rightarrow\gamma gg,\;gg\rightarrow f_2$.
The coupling between $f_2$ and two gluons is written as[8]
\begin{eqnarray}
G^{ab}_{\alpha\beta,\lambda_2}(x_1,x_2)=<f_{gg\lambda_2}|T\{A^a_{\alpha}(x_1)A^b_{\beta}(x_2)\}|0>
=\delta_{ab}e^{{i\over2}p_f(x_1x_2)}G(0)\sum_{m_1 m_2}c^{2\lambda_2}_{1m_1 1m_2}e^{*m_1}_{\alpha}e^{*m_2}_{\beta},
\end{eqnarray} 
where G(0) is taken as a parameter. Using Eq.(6), the helicity amplitudes of $J/\psi\rightarrow\gamma f_2$ are presented
in Ref.[8]. Replacing $m_c$ by $m_b$ in Eqs.(3,4,11) of Ref.[8], the helicity amplitudes of 
$\Upsilon(1S)\rightarrow\gamma f_2$ are obtained 
\begin{eqnarray}
\lefteqn{T_0=-{2\over\sqrt{6}}(A_2+p^2 A_1),}\nonumber\\
&&T_1=-{\sqrt{2}\over m_J}(E A_2+m_f p^2 A_3),\nonumber\\
&&T_2=-2A_2,\\
&&E={1\over2m_f}(m^2_{\Upsilon}+m^2_f),\;\;p={1\over2m_f}(m^2_{\Upsilon}-m^2_f),
\end{eqnarray}
where 
\begin{eqnarray}
\lefteqn{A_1=-a\frac{2m^2_f-m_J(m_{\Upsilon}-2m_b)}{m_b m_{\Upsilon}[m^2_b+{1\over4}(m^2_{\Upsilon}-2m^2_f)]}
,}\nonumber\\
&&A_2=-a{1\over m_b}\{{m^2_f\over m_{\Upsilon}}-m_{\Upsilon}+2m_b\},\nonumber\\
&&A_3=-a\frac{m^2_f-{1\over2}(m_{\Upsilon}-2m_b)^2}{m_b m_{\Upsilon}[m^2_b+{1\over4}(m^2_{\Upsilon}-2m^2_f)]},\nonumber\\
&&a={16\pi\over 3\sqrt{3}}\alpha_s(m_b) G(0)\psi_J(0){\sqrt{m_{\Upsilon}}\over m^2_b},
\end{eqnarray}
where $\psi_{\Upsilon}(0)$ is the wave functions of $\Upsilon$ at origin.
The decay width of $\Upsilon\rightarrow\gamma f_2$ is derived as
\begin{equation}
\Gamma(\Upsilon\rightarrow\gamma f_2)=\frac{32\pi\alpha}{81}sin^2\phi\alpha_s^2(m_b)G^2(0)\psi^2_{\Upsilon}(0){1\over m^4_b}
(1-{m^2_f\over m^2_{\Upsilon}})\{T^2_0+T^2_1+T^2_2\}.
\end{equation}
The ratios of the helicity amplitudes are defined as
\begin{equation}
x={T_1\over T_0},\;\;y={T_2\over T_0}.
\end{equation}
The expressions of these quantities for $J/\psi\rightarrow\gamma f_2$ can be found from Ref.[8].

The wave functions of $\Upsilon$ or $J/\psi$ at the origin are related to their rates of decaying to $ee^+$
\begin{equation}
\frac{\psi^2_\Upsilon(0)}{\psi^2_J(0)}=4\frac{\Gamma_{\Upsilon\rightarrow ee^+}}{\Gamma_{J/\psi\rightarrow ee^+}}
\frac{m^2_\Upsilon}{m^2_{J/\psi}}.
\end{equation}
The parameters $sin^2\phi G^2(0)$ are canceled in the ratio 
\[R=\frac{B(\Upsilon\rightarrow\gamma f_2)}
{B(J/\psi\rightarrow\gamma f_2)}\]
Taking
\(\alpha_s(m_c)=0.3\), \(\alpha_s(m_b)=0.18\)[6], and \(m_c=1.29GeV\)(experimental value is 
\(m_c=1.27^{+0.07}_{-0.11}GeV\)[4]),  
\(m_b=(5.04\pm0.075\pm0.04)GeV\)[4] is obtained
\begin{equation}
R=0.071(1\pm0.17)
\end{equation}
which agrees with experimental data[4].

The ratios of the helicity amplitudes are obtained 
\begin{equation}
x^2=0.058,\;\;y^2=5.9\times10^{-3}.
\end{equation}
They are consistent with experimental values[1] 
\begin{equation}
x^2=0.00^{+0.02+0.01}_{-0.00-0.00},\;\;y^2=0.09^{+0.08+0.04}_{-0.07-0.03}.
\end{equation}

Eqs.(7,9,10) show that the approach[8] used to study the decay $\Upsilon\rightarrow\gamma f_2$
leads to strong suppression by the mass of b-quark. On the other hand, eq.(14) shows that
this approach leads to 
\begin{equation}
A_2\sim 0
\end{equation}
and very small $T_{1,2}$. Therefore, the amplitude $T_0$ makes dominant contribution to 
the decay rate of $\Upsilon\rightarrow\gamma f_2$. Because of Eq.(16) 
\begin{equation}
\Gamma(\Upsilon\rightarrow\gamma f_2)\propto p^2,
\end{equation}
Eq.(17) leads to a strong enhancement for the decay rate. The $T_0$ dominance has been found in Ref.[6] and 
\(R\sim 0.059\) is obtained. In Ref.[6] \(m_c=1.5GeV\)
is taken. The value used in this study is consistent with the experimental data[4] and the amplitudes are
sensitive to the value of $m_c$.
Therefore, there is competition between the suppression and
the enhancement in the decay $\Upsilon\rightarrow\gamma f_2$. 
In QCD $J/\psi,\;\Upsilon\rightarrow light\;
hadrons$ are described as $J/\psi,\Upsilon\rightarrow 3g$ whose decay width is proportional to $\alpha_s^3 m_V$,
where $m_V$ is the mass of $J/\psi, \Upsilon$ respectively. Putting these factors together, the ratio is expressed as
\begin{eqnarray}
\lefteqn{R=\frac{B(\Upsilon\rightarrow\gamma f_2)}{B(J/\psi\rightarrow\gamma f_2)}=
\frac{\Gamma(\Upsilon\rightarrow\gamma f_2)}{\Gamma(J/\psi\rightarrow\gamma f_2)}\frac{\Gamma(J/\psi\rightarrow lh)}
{\Gamma(\Upsilon\rightarrow lh)}\frac{B(\Upsilon\rightarrow lh)}{B(J/\psi\rightarrow lh)} }\nonumber \\
&&=1.06\frac{\alpha_s(m_c)}{\alpha_s(m_b)}\frac{p^4_\Upsilon}{p^4_J}
\frac{m_J m^6_c}{m_\Upsilon m^6_b}\frac{[m^2_c+{1\over4}(m^2_J-2m^2_f)]^2}
{[m^2_b+{1\over4}(m^2_\Upsilon-2m^2_f)]^2}\frac{(1-{m^2_f\over m^2_\Upsilon})}{(1-{m^2_f\over m^2_\Upsilon})}\nonumber \\
&&\frac{\{2m^2_f-m_\Upsilon(m_\Upsilon-2m_b)\}^2}{\{2m^2_f-m_J(m_J-2m_c)\}^2
+6{m^2_f\over m^2_J}\{m^2_f-{1\over2}(m_J-2m_c)^2\}^2},
\end{eqnarray}
where  
\begin{equation}
p_J={m^2_J\over2m_f}(1-{m^2_f\over m^2_J}).
\end{equation}
The competition between the suppression and
the enhancement in the decay $\Upsilon\rightarrow\gamma f_2$
makes the dependence of 
$\frac{B(\Upsilon\rightarrow\gamma f_2)}{B(J/\psi\rightarrow\gamma f_2)}$ on quark masses much weaker than the ratio
$\frac{B(\Upsilon\rightarrow\gamma \eta'(\eta)}{B(J/\psi\rightarrow\gamma\eta'(\eta))}$[5]. 

In summary, the approach[4] in which $f_2(1270)$ is strongly coupled to two gluons leads to very small
ratios of helicity amplitudes, x and y, and not small branching ratio of $\Upsilon\rightarrow\gamma f_2$

\end{document}